\documentstyle[prb,aps,epsfig,twocolumn]{revtex} 
 
\begin{document}

\title{Influence of uncorrelated overlayers on the magnetism in thin
  itinerant-electron films} 
 
\author{J.H. Wu\cite{jhwu}, T. Herrmann, and W. Nolting}  
 
\address{Lehrstuhl Festk\"orpertheorie, Institut f\"ur Physik,
Humboldt-Universit\"at zu Berlin,  
Invalidenstr. 110, 10115 Berlin, Germany} 
 
\maketitle 
 
\begin{abstract} 
The influence of uncorrelated (nonmagnetic) overlayers on the magnetic
properties of thin itinerant-electron films is investigated within the
single-band Hubbard model. The Coulomb correlation between the electrons 
in the ferromagnetic layers is treated by using the spectral density
approach (SDA). It is found that the presence of nonmagnetic layers has
a strong effect on the magnetic properties of  thin films. The Curie
temperatures of very thin films are modified by the uncorrelated
overlayers. The quasiparticle density of states is used to analyze the
results. In addition, the coupling between the ferromagnetic layers and
the nonmagnetic layers is discussed in detail. The coupling  depends on
the band occupation of the nonmagnetic layers, while it is almost
independent of the number of the nonmagnetic layers. The induced
polarization in the nonmagnetic layers shows a long-range decreasing
oscillatory behavior and it depends on the coupling between
ferromagnetic and nonmagnetic layers.\\ 
\end{abstract} 

\pacs{75.70.Ak, 75.70.Cn, 71.10.Lp}

\section{INTRODUCTION} 

In the past few years, there has been growing interest in the magnetic 
behavior of thin metallic films. This attention originates both from 
fundamental physics and from applications. The reduced symmetry and
the lower coordination numbers at the surface offer the possibility of
inducing new and exotic phenomena, such as perpendicular 
anisotropy in ultrathin films\cite{ASB90}, interlayer coupling\cite{GSP+86}
and giant magnetoresistance(GMR)\cite{BBF+88} in ferromagnetic metal/nonmagnetic 
metal (FM/NM) multilayers. In experiment, it was shown that ultrathin
transition metal films can display long range ferromagnetic order from a
monolayer on\cite{DTP+89}. It is well known, that due to the 
Mermin-Wagner theorem\cite{MW66}, an effectively two-dimensional 
spin-isotropic system cannot display 
long-range magnetic order at any finite temperature. However,  
in real ultrathin films, there always exists an anisotropy which allows 
almost two-dimensional magnetism to occur\cite{BM88}. 

In experiment, the magnetic thin films are normally grown on a nonmagnetic 
substrate. The effects of the presence of nonmagnetic overlayers on the 
magnetic properties of these thin films are widely studied both in
experiment\cite{WWH+92,DKW+99,WBB+95,DB99,WKP+90,EWL+93,CKE+99,GB84,HBF+88,CHC+90,HHS+91,OH92,GCJ93,CVR+93}
and theory\cite{ZKW+96,RVR+94,FFO85,GR86,BWD88,SHL+92}. It has been shown that
the interfaces between 
magnetic and nonmagnetic layers play an important role with respect to
the magnetic properties of multilayers systems\cite{WWH+92}. The composition
of the interface between FM and NM has a strong effect on the magnetic
properties of the FM films\cite{DKW+99}. The critical Co film thickness for the
reorientation transition of the magnetization has been observed to  
shift from 3--4 ML up to 18 ML by the presence of carbon at the 
interface\cite{DKW+99}. In addition, very low coverage of nonmagnetic
materials on top of a magnetic layers can have a strong effect on the
direction of the magnetization\cite{WBB+95,DB99,WKP+90,EWL+93}. For
example, an almost 
90$^\circ$ rotation from in-plane direction to out-of-plane direction of
the magnetization is induced by 0.03 monolayer(ML) Cu on a 7 ML thick Co
film on a stepped Cu(001) surface\cite{WBB+95}. The magnetic properties of
ultrathin films do not change monotonically with the overlayer
thickness in some systems. The magnetic anisotropy of thin FM film is observed to rotate
from in-plane direction to out-of-plane direction with very thin NM
overlayers, while it will return to in-plane direction when the NM
overlayers become thicker\cite{WBB+95,DB99,WKP+90,EWL+93}. First
principle calculations 
for a Co ML on Cu(111) show a switch from the in-plane anisotropy of
the uncovered Co monolayer to perpendicular anisotropy only for a 1 ML
thick Cu overlayer, while 2 ML Cu produce a slight in-plane anisotropy
again\cite{ZKW+96}.

The induced polarization in the NM layers has been investigated 
experimentally in  various systems, such as Pt/Co\cite{WWH+92}, Pd/Ni\cite{GB84}, 
Pd/Fe\cite{DB99,HBF+88,CHC+90,RVR+94}, Pd/Co\cite{WWH+92,HHS+91}, and
Cu/Co\cite{OH92,GCJ93,CVR+93}.  
The direction of the induced polarization has been found to be 
different for different systems. In 
Pd/Fe films, the Pd near the interface is  ferromagnetically
coupled with the Fe film\cite{DB99}. The induced moment of the  Pd has been  
shown to enhance the Curie temperature $T_C$ of the film\cite{DB99}. 
On the other hand, in Pt/Co
systems  with thicknesses up to 1.5 ML  the Pt layer is 
negatively spin polarized (antiferromagnetic coupling)\cite{GB84}. 
In Cu/Co systems the spin
polarization of the Cu layers has been shown to oscillate as a function of the
Cu thickness. 
This oscillation is regarded as the origin of the oscillation of the 
interlayer magnetic coupling in Cu/Co multilayers \cite{OH92,GCJ93,CVR+93}.   
In theory\cite{RVR+94,FFO85}, based on  first principle calculations, the
polarization has been found to be positive both in Pd/Fe\cite{RVR+94} and 
NM/TM\cite{FFO85} (NM denotes noble metal such as Au, Ag and Cu, TM: Fe,
Cr) systems. Systematic calculations show that Fe, Co and Ni
overlayers on Pd favour a ferromagnetic configuration, whereas V, Cr and
Mn overlayers lead to an antiferromagnetic superstructure\cite{BWD88}. 

Clearly,  nonmagnetic overlayers play a very important role with respect to
the magnetic properties of  thin metallic films. 
Theoretically, the influence of nonmagnetic overlayers on transition metal
films is  mostly investigated within first principle 
calculations\cite{RVR+94,FFO85,GR86,BWD88,ZKW+96}. However, these calculations   
can give the zero temperature properties of the films only. 
Certain idealized model systems have proved to be a good starting point for 
investigating the magnetic behavior at finite temperatures. 
In order to investigate the temperature dependence of the magnetization, the 
Ising model was adopted in Ref.\onlinecite{SHL+92}. However, in transition
metals, the magnetically active electrons are itinerant. It is by no
means clear to what extent the results obtained by localized spin models
are applicable to transition metal films. In addition, the quantum
interference of Bloch waves in the NM layers has been applied to study
the interlayer magnetic coupling in the FM/NM
multilayers\cite{BRU95}. Within this theory, only the properties of the NM
layers are considered, while the properties of the FM layers are
neglected completely.  In a theory
of interlayer magnetic coupling in the FM/NM multilayers, Edwards 
{\em et al.} obtained the difference of exchange coupling between
two FM layers for difference band occupation in NM
layers\cite{EMM+91}. Their results corresponds to a Hartree-Fock
treatment of a Hubbard-like one-band tight-binding model with on-site
interaction $U=0$ in the NM and with $U=\infty$ in the FM. However, they 
didn't calculate the magnetization of the FM and NM, and the effect of
the NM on the properties of the FM was neglected.

The aim of the present paper is to investigate the influence of NM
overlayers on the magnetism of itinerant-electron thin films within the Hubbard
model\cite{HUB63}. The influence of the coupling between the FM and NM
layers on both the magnetic properties of the FM layers and the induced
magnetization in the NM layers will be investigated in detail.

The Hubbard model was originally introduced to explain band magnetism in
transition metals and has become a standard model to study the essential
physics of strongly correlated electron systems over years, such as
spontaneous magnetism, metal-insulator transition and high-temperature
superconductivity. It incorporates in the simplest way the interplay of
the kinetic energy, the Coulomb interaction, the Pauli principle and the
lattice geometry. In systems with reduced translational symmetry, 
the Hubbard model has been
successfully applied to describe the temperature-driven reorientation
transition of magnetization in itinerant-electron films\cite{HPN98},
metal-insulator transition in thin films\cite{PN99}, the surface
magnetism\cite{PN971} and low dimensional magnetism\cite{HN99}. The model,
though in principle rather simple, nevertheless, provokes a non-trivial
many-body problem, that could be solved up to now only for some special
cases\cite{LW68,EME79,HAL81,MS92,VOI95,MV89,MUL89,VOL93,GKK+96}. In two and
three dimensions, one still has to resort to approximate treatments.
Due to the reduced translation symmetry in thin films even more complications
are introduced. Recently a generalization of the spectral density
approach (SDA)\cite{NB89,HN97} has been applied to study the magnetism of thin
metallic films and surfaces \cite{HPN98,HN99,PN96,HER99}. The SDA, which
reproduces  the exact results of  Harris and Lange
\cite{HL67} that concern the general shape of the spectral density 
in the strong-coupling limit ($U\gg W$, $U$: on-site Coulomb interaction, 
$W$: bandwidth of the Bloch density of states), 
leads to rather convincing results
concerning the magnetic properties of the Hubbard
model\cite{HN99,NB89,HN97,PN96,HER99}. 
By comparison with different approximation schemes for the Hubbard model 
as well as numerically exact QMC calculations in the limit of infinite
dimensions it has been shown \cite{PHWN98} that the
correct inclusion of the exact results of Harris and Lange \cite{HL67}
in the strong coupling regime is vital for a 
reasonable description of the magnetic behavior of
the Hubbard model, especially at finite temperatures.

The structure of the paper is as follows. First, the Hamiltonian of our
model is proposed. Then the SDA for the Hubbard film is described in a simple
way. In section IV we show the results of the numerical evaluation of the 
theory and discuss the magnetic behavior of the film system in terms of the 
layer and temperature dependent magnetizations and the quasiparticle density of
states.  Finally, a summary will be given.
  
\section{Hamiltonian of the model} 

In this paper, we will concentrate on the essence of the effect of
NM overlayers on the magnetic properties of itinerant-electron thin
films, as well as the influence of the FM layers on the NM layers. The
structure discussed here is a NM/FM/NM  sandwich structure. 
We study the symmetric situation where the numbers of the
NM layers both above and below the FM layers are assumed to be equal.

The description of this film geometry requires some care. Each lattice 
vector of the film is decomposed into two parts: 
\begin{equation} 
\label{e1} 
{\bbox R}_{i\alpha}={\bbox R}_i+{\bbox r}_\alpha
\end{equation} 
${\bbox R}_i$ denotes a lattice vector of the two-dimensional
Bravais lattice of the surface layer with $N$ sites. 
To each lattice point $\bbox{R}_i$ a $d$-atom basis
${\bbox r}_\alpha$ ($\alpha=1,2,\cdots,d$) is associated,
referring to the $d=2d_{NM}+d_{FM}$  layers of the
film. Here, $d_{NM}$ denotes the
thickness of the NM layers and $d_{FM}$ the thickness of the FM layers.
Within each layer we assume translational invariance. Then a
Fourier transformation with respect to the underlying two-dimensional (surface)
Bravais lattice can be applied.
        
The considered model Hamiltonian reads as follows:
\begin{equation} 
\label{e2} 
{\cal H}=\sum_{i,j,\alpha,\beta,\sigma}(T^{\alpha\beta}_{ij}-
\mu\delta^{\alpha\beta}_{ij}) c^+_{i\alpha\sigma}c_{j\beta\sigma}+
\frac{U}{2}\sum_{i,\alpha,\sigma} V(\alpha)
n_{i\alpha\sigma}n_{i\alpha-\sigma}, 
\end{equation} 
where $c^+_{i\alpha\sigma}(c_{i\alpha\sigma})$ stands for the creation
(annihilation) operator of an electron with spin $\sigma$ at the lattice site
$\bbox{R}_{i\alpha}$, $n_{i\alpha\sigma}=c^+_{i\alpha\sigma}c_{i\alpha\sigma}$ is the
number operator, and $T^{\alpha\beta}_{ij}$ denotes the hopping-matrix 
element between the lattice sites $\bbox{R}_{i\alpha}$ and $\bbox{R}_{j\beta}$. 
The hopping-matrix element between nearest-neighbour sites 
is set to $-t_{FM}$ ($-t_{NM}$) in the FM (NM) layers and to
$-t_{NF}$ between FM and NM layers. 
$U$ denotes the on-site Coulomb matrix element and $\mu$ is
the chemical potential. 
The Coulomb interaction between the electrons is only considered in the FM
layers. Therefore we choose for $V(\alpha)$: 
\begin{equation} 
\label{e3} 
V(\alpha)=\left \{
\begin{array}{ll}
0,~~~&\alpha\leq d_{NM} \mbox{ or } \alpha > d_{NM}+d_{FM},\\
1,~~~&  d_{NM} < \alpha \leq d_{NM}+d_{FM}.
\end{array}
\right.
\end{equation} 

In the following, all quantities related to the NM layers will be
labelled by a subscript $NM$, those related to the FM layers by a
subscript $FM$.

\section{Spectral-density approach to the Hubbard film} 

Recently a generalization of SDA has been proposed to deal with the
modifications due to reduced translational
symmetry. In 
the following we give only a brief derivation of the SDA solution and
refer the reader to previous papers for a detailed
discussion\cite{HPN98,HN99,HN97,PN96,HER99}.

The basic quantity to be calculated is the retarded single-electron Green
function 
\begin{equation}
\label{e4a}
G^{\alpha\beta}_{ij\sigma}(E)=\langle\langle c_{i\alpha\sigma};
c^+_{j\beta\sigma}\rangle\rangle_E.
\end{equation} 
All relevant information about the
system can be obtained from the Green function. For example, the spin- and
layer-dependent quasiparticle density of states (QDOS) is determined by
the diagonal part of $G^{\alpha\beta}_{ij\sigma}(E)$:  
\begin{equation} 
\label{e4} 
\rho_{\alpha\sigma}(E)=-\frac{1}{\pi}\mbox{Im}G^{\alpha\alpha}_{ii\sigma}(E-\mu).
\end{equation} 

The band occupations $n_{\alpha\sigma}$ are given by
\begin{equation} 
\label{e5} 
n_{\alpha\sigma}=\int^{+\infty}_{-\infty} dE f_-(E) \rho_{\alpha\sigma}(E),
\end{equation} 
where $f_-(E)$ is the Fermi function. Ferromagnetism is indicated by a
spin-asymmetry in the band occupations $n_{\alpha\sigma}$ leading to
non-zero layer magnetizations
$m_\alpha=n_{\alpha\uparrow}-n_{\alpha\downarrow}$. The band
occupation in each layer is given by $n_{\alpha}=n_{\alpha\uparrow}+
n_{\alpha\downarrow}$.

The equation of motion for the single-electron Green function reads:
\begin{equation} 
\label{e6} 
\sum_{l,\gamma}[(E+\mu)\delta^{\alpha\gamma}_{il}-T^{\alpha\gamma}_{il}-
V(\gamma)\Sigma_{il\sigma}^{\alpha\gamma}(E)] G_{lj\sigma}^{\gamma\beta}(E) =
\hbar \delta^{\alpha\beta}_{ij}.
\end{equation} 
Here we have introduced the electronic self-energy
$\Sigma_{ij\sigma}^{\alpha\beta}(E)$ which incorporates all effects of
electron correlations. The self-energy will automatically vanish
within the NM layers due to $V(\alpha)=0$. The local approximation for
self-energy in the FM layers is adopted, which has been tested recently
for the case of reduced translational symmetry\cite{PN972}. Because the
translational invariance is assumed within each layer, we then have
$\Sigma^{\alpha\beta}_{ij\sigma}(E)=\delta^{\alpha\beta}_{ij}
\Sigma^{\alpha}_{\sigma}(E)$. 

The key point of the SDA is to find a reasonable ansatz for the self-energy
in FM layers. Guided by the exactly solvable atomic limit of vanishing
hopping ($t_{FM}=0$) and by the findings of Harris and Lange \cite{HL67} in the
strong-coupling limit ($U/t_{FM}>>1$), a one-pole ansatz for the self-energy 
$\Sigma_{\sigma}^{\alpha}(E)$ can be motivated\cite{PN96}:
\begin{equation}
\label{e7}
\Sigma^{\alpha}_{\sigma}(E)=g^{\alpha}_{1\sigma}
\frac{E-g^{\alpha}_{2\sigma}}{E-g^{\alpha}_{3\sigma}}
\end{equation}
where the spin- and layer-dependent parameters $g^{\alpha}_{1\sigma}$,
$g^{\alpha}_{2\sigma}$ and $g^{\alpha}_{3\sigma}$ are fixed by
exploiting the equality between two alternative but exact
representations for the moments of the layer-dependent QDOS:
\begin{equation}
\label{e8}
\begin{array}{ll}
M_{ij\sigma}^{(m)\alpha\beta}&=-\frac{1}{\pi\hbar}\mbox{Im}
\int^{\infty}_{-\infty} dE (E)^m G^{\alpha\beta}_{ij\sigma}(E)\\
~~&~~\\
&=\langle  [\underbrace{\left [\cdots[c_{i\alpha\sigma},\cal{H}]_-\cdots,
 \cal{H}\right ]_-}_{m-\mbox{times}},c_{j\beta\sigma}^+ ]_+\rangle .  
\end{array}
\end{equation}  
Here $\langle\cdots\rangle$ denotes the grand-canonical average and
$\left[\cdots,\cdots\right]_{-(+)}$ is the commutator
(anticommutator). It has been shown\cite{PHWN98} that an inclusion of the
first four moments of QDOS ($m=$0--3) is vital for a proper 
description of ferromagnetism in the Hubbard model, especially for finite
temperatures. Further, the first
four moments represent a necessary condition \cite{PHWN98}
to be consistent with the
strong coupling results of Harris and Lange. 
Taking into the first four moments
to fix the three parameters $g^{\alpha}_{1\sigma}$,
$g^{\alpha}_{2\sigma}$ and $g^{\alpha}_{3\sigma}$ in (\ref{e7}), 
one obtains the
following self-energy \cite{PN96,HER99,PHWN98}:
\begin{equation}
\label{e9}
\Sigma_{\sigma}^{\alpha}(E)=Un_{\alpha-\sigma}\displaystyle
\frac{E+\mu-B_{\alpha-\sigma}}{E+\mu-B_{\alpha-\sigma}-U(1-n_{\alpha-\sigma})}.
\end{equation}
The self-energy depends on the spin-dependent occupation numbers
$n_{\alpha\sigma}$ and the so-called band-shift $B_{\alpha\sigma}$ that
consists of higher correlation functions:
\begin{equation}
\label{e10}
B_{\alpha\sigma}=T^{\alpha\alpha}_{ii}+\displaystyle
\frac{1}{n_{\alpha\sigma}(1-n_{\alpha\sigma})}\sum_{j\beta}^{j\beta\neq 
  i\alpha} T^{\alpha\beta}_{ij}\langle
c^+_{i\alpha\sigma}c_{j\beta\sigma}(2n_{i\alpha-\sigma}-1)\rangle. 
\end{equation}
A spin-dependent shift of the band center of gravity $n_{\alpha-\sigma}
B_{\alpha-\sigma}$ $\left ((1-n_{\alpha-\sigma})B_{\alpha-\sigma}\right)$ may
generate and stabilize ferromagnetic solutions. Although
$B_{\alpha\sigma}$ consists of higher correlation functions, it can be
expressed exactly via $\rho_{\alpha\sigma}$ and 
$\Sigma_\sigma^\alpha(E)$\cite{PN96,HER99,PHWN98}:
\begin{equation}
\label{e11}
\begin{array}{ll}
B_{\alpha\sigma}&=T^{\alpha\alpha}_{ii}+\displaystyle
\frac{1}{n_{\alpha\sigma}(1-n_{\alpha\sigma})}\frac{1}{\hbar}
\int^{+\infty}_{-\infty} dE f_-(E)\\
~~&~~\\
&\times\left(\frac{2}{U}\Sigma^\alpha_\sigma(E-\mu)-1\right)
\left[E-\Sigma_\sigma^\alpha(E-\mu)-T^{\alpha\alpha}_{ii}\right]
\rho_{\alpha\sigma}(E).
\end{array}
\end{equation}
Now a closed set of equations is established via Eqs.(\ref{e5}),
(\ref{e6}), (\ref{e9}) and (\ref{e11}), which can be solved self-consistently.

\section{Results and discussion}

In our calculations, a fcc(100) geometry is assumed for both the FM and
the NM layers. We consider a uniform hopping $t_{FM}=t_{NM}=t_{NF}=0.25$eV
between nearest neighbour sites.
The band occupation of the FM layers is set to $n_{FM}=1.4$ for all 
calculations. The band occupation in the NM layers is denoted by $n_{NM}$.
By adjusting the on-site hopping integrals $T_{ii}^{\alpha\alpha}$ 
we explicitly exclude charge transfer within the FM and the NM layers.
Further, we keep the on-site Coulomb interaction
in the  FM layers fixed at $U=12$eV, which is three times the bandwidth of the
three-dimensional fcc lattice and clearly refers to the strong-coupling
regime. In the following we will refer to the considered structure as
$d_{NM}/d_{FM}/d_{NM}$. 
\begin{figure} 
\centerline{\epsfig{figure=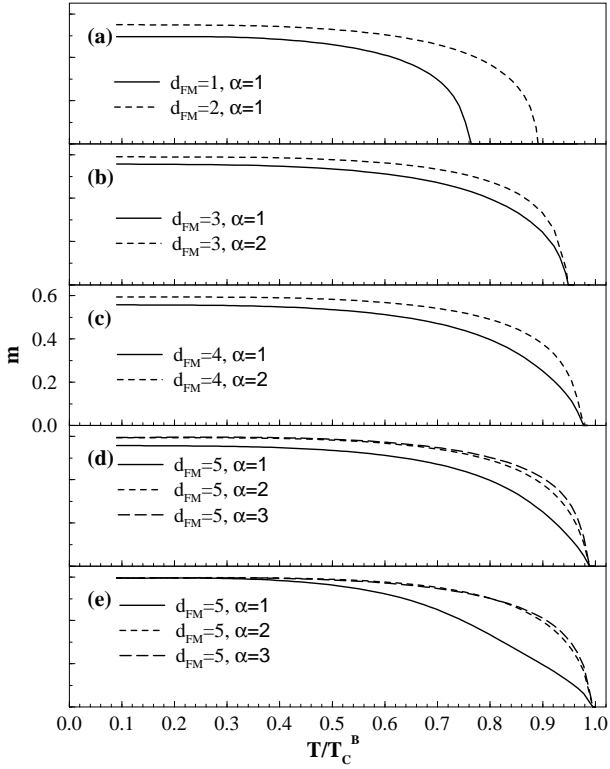,width=0.9\linewidth,angle=0}}
\vspace*{0.5cm}
\caption{The layer-dependent magnetization $m_\alpha$ of the FM layers 
  for a  $d_{NM}/d_{FM}/d_{NM}$
  sandwich structure as a function of  temperature
  for different numbers $d_{FM}$ of  FM layers. 
  (a) $d_{FM}=1$ and 2, (b),
  (c), (d) refer $d_{FM}=3,4$ and 5, respectively. For (a)--(d),
  the number of NM layers is $d_{NM}=4$.
  (e) $d_{FM}=5$ and
  $d_{NM}=0$ (free standing  FM film). $\alpha=1$ denotes the
  interface FM layer, and $\alpha=2,3$ the inner FM layers.  
  Other parameters are:
  $n_{FM}=1.4$, $n_{NM}=0.8$, $T_C^B$ refers to the bulk Curie temperature.}  
\end{figure} 

In Fig.~1, the layer-dependent magnetizations are plotted as a function of 
temperature for different numbers of magnetic layers. The magnetizations
are strongly layer-dependent. Without  NM overlayers (Fig.~1e; 0/5/0
structure), the magnetization in each layer is fully polarized at very
low temperatures. The magnetization curves of the inner layers show the
usual Brillouin-type behavior. The surface magnetization, however, shows
a very different behavior: It depends almost linearly on temperature in
the temperature range $T/T_C^B=0.7$--$0.9$. Compared to the inner
layers,  the surface magnetization decreases significantly faster as a
function of temperature and shows a tendency to a reduced Curie
temperature. However, due to the coupling between the surface and the
inner layers that is induced by the electron hopping, a unique Curie
temperature for the whole film is obtained\cite{HPN98}. When the NM
overlayers are taken into account (see Fig.~1a,1b,1c,1d; 4/$d_{FM}$/4
structure), the interface magnetization of the FM layers and its
temperature behavior are strongly affected by the interaction between
the FM and NM layers which is induced by the electron hopping at the
interface. The interface is no longer fully polarized at low
temperatures and decreases more roundly as a function of temperature
than in the case  without  NM overlayers. The linear dependence of the
interface magnetization in the range of $T/T_C^B=0.7$--$0.9$ has
disappeared. The magnetization of the inner FM layers is only weakly
affected by the presence of the NM layers (see Fig.~1d). 
\begin{figure} 
\centerline{\epsfig{figure=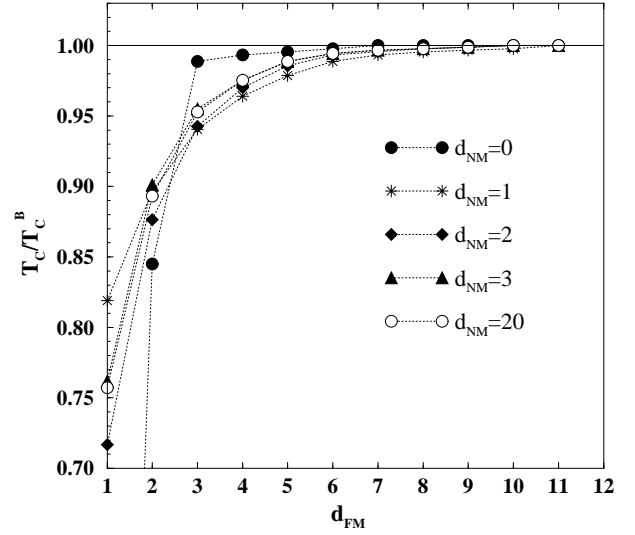,width=0.9\linewidth,angle=0}}
\vspace*{0.5cm}
\caption{The Curie temperature $T_C$ as a function of the number $d_{FM}$
  of FM
  layers for different numbers $d_{NM}$ of NM layers. $d_{NM}=0$ refers to the 
  free standing FM film. The parameters are: $n_{FM}=1.4$, $n_{NM}=0.8$.} 
\end{figure} 


From Fig.~1, one can also read off that the Curie temperature $T_C$
increases gradually as a function of the number of the FM layers
$d_{FM}$ and reaches its bulk value for $d_{FM}=8$--$10$. The influence
of the NM layers on the Curie temperature is analyzed in more detail in
Fig.~2 where $T_C(d_{FM})$ is shown for different numbers of the NM
layers. As can be seen in Fig.~2, the behavior of $T_C$ as a function of
$d_{FM}$ is quite different for $d_{NM}=0$ and $d_{NM}\neq 0$. In the
case without NM overlayers ($d_{NM}=0$) $T_C$ increases very steeply as
a function of $d_{FM}$ and saturates already for $d_{FM}=$3--5\cite{HPN98}.
The influence of the NM layers is most important for very thin magnetic
films $d_{FM}<6$. For $d_{FM}=3,4$ the presence of only one NM overlayer
leads to  a reduction of the Curie temperature of about 5\% compared to
its free FM film value. It is interesting to note that there are only
minor changes in $T_C$ when the number of NM layers is further
increased. Note that for $d_{FM}\geq 3$, the reduction of the $T_C$ is
strongest for just one NM layer. As $d_{NM}$ is further increased, the
reduction of the $T_C$ becomes slightly decreasing again. $T_C$ as a
function of $d_{NM}$ saturates already for $d_{NM}=3$. This indicates
that the influence of a very thin NM topping on the magnetic properties of
the FM films is stronger than for thick NM overlayers. In
experiment\cite{WBB+95,DB99,WKP+90,EWL+93}, it has also been observed
that very thin NM 
layers have a stronger effect on the magnetic properties than thick
overlayers. For example, the direction of the magnetization of a Co film
shows a rotation from the in-plane direction to the out-of-plane
direction induced only by very thin coverage of Cu, while for thick
coverage of Cu, the direction of magnetization of the FM films will turn 
back to in-plane direction\cite{WBB+95,DB99,WKP+90,EWL+93}. First
principle calculations 
for a Co ML on Cu(111) predict a switch from the in-plane anisotropy of
the uncovered Co monolayer to perpendicular anisotropy only for a 1 ML
thick Cu overlayer, while 2 ML Cu produces a slight in-plane anisotropy
again\cite{ZKW+96}. 

\begin{figure} 
\centerline{\epsfig{figure=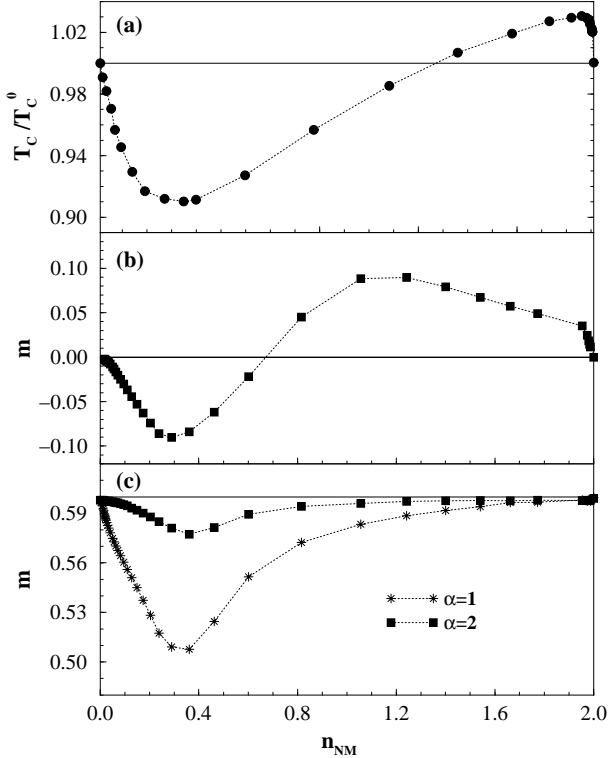,width=0.9\linewidth,angle=0}}
\vspace*{0.5cm}
\caption{The Curie temperature $T_C$ and magnetizations $m_\alpha$ 
  of a $d_{NM}/d_{FM}/d_{NM}$=1/3/1 sandwich structure as functions of the
  band occupation $n_{NM}$ of the NM layers. (a) $T_C$, (b) the magnetization of
  the NM layer, and (c) the magnetizations of the FM layers. $\alpha=1$
  indicates the interface FM layer. The fully polarized magnetizations
  of the FM layers are 0.6.
   $T_C^0$ denotes the Curie temperature of the free FM
  film. Other parameters are: $n_{FM}=1.4$. For (b) and (c) the
  temperature is $T=0.1T_C^0$.} 
\end{figure} 

In order to investigate the influence of the NM overlayers on the magnetism 
of the FM films in more detail, we have calculated the Curie temperature
as a function of the band occupation $n_{NM}$ of the NM layer for a 1/3/1
snadwich structure (Fig.~3a). 
$T_C$ does not change monotonically as a function of $n_{NM}$. For very
small $n_{NM}$ it decreases rapidly from the free FM film value $T_C^0$
($d_{NM}=0$) to a minimum at $n_{NM}=$0.35--0.4. Then it increases from
the minimum to a maximum (at $n_{NM}\approx 1.97$) which lies above
$T_C^0$. Finally, it drops quickly to $T_C^0$ at $n_{NM}=2.0$. 
The two limiting cases ($T_C=T_C^0$ at $n_{NM}=0$ and $n_{NM}=2.0$)
are easy to understand. In these two situations, the NM bands are either
empty or fully occupied and, therefore, have no influence on the
properties of the FM film. To understand the behavior of $T_C$ as a
function of $n_{NM}$, we have also calculated the magnetizations of the
NM layer and the FM layers as a function of $n_{NM}$ at low temperature
in Fig.~3b and Fig.~3c. The magnetization of the NM layer shows a
similar dependence on $n_{NM}$ as the Curie temperature (see Fig.~3b). 
Since the magnetization of the FM layers is always positive,
the sign of magnetization of the NM layer represents the coupling
between FM and NM at the interface. A positive sign corresponds to 
ferromagnetic coupling, a negative sign to antiferromagnetic coupling. 
The results indicate that $T_C$ is strongly affected by the coupling of
FM and NM. From Fig.~3c, one can see that the coupling of FM and NM has
also a strong effect on the magnetizations in the FM layers, especially
on the interface magnetization at low temperature.

\begin{figure} 
\centerline{\epsfig{figure=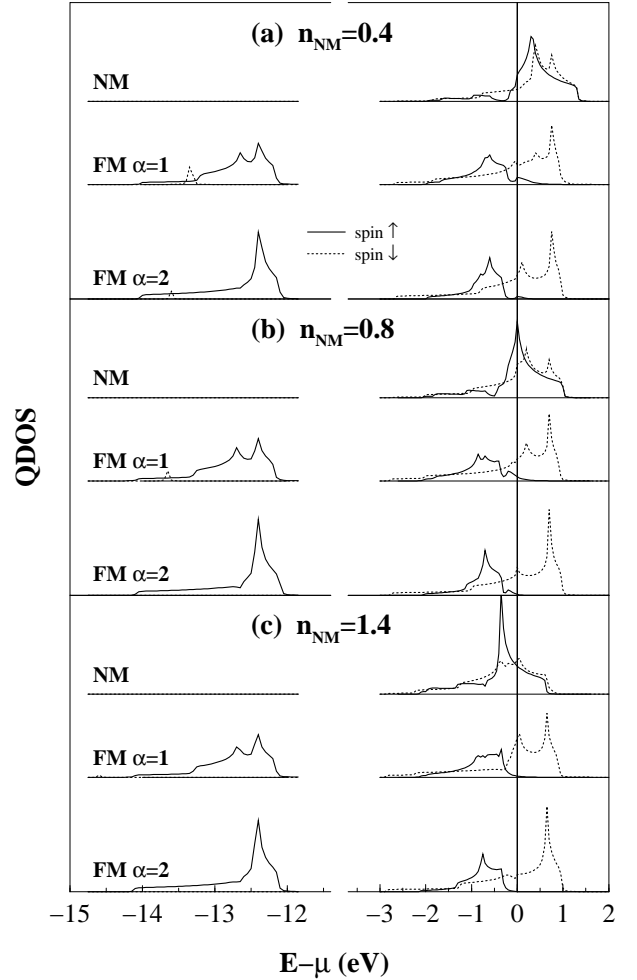,width=0.9\linewidth,angle=0}}
\vspace*{0.5cm}
\caption{Layer-dependent QDOS of a 1/3/1 sandwich structure for different
  $n_{NM}$ at low   temperature ($T=0.1 T_C^0$). (a)
  $n_{NM}=0.4$, (b) $n_{NM}=0.8$ and (c) $n_{NM}=1.4$.}  
\end{figure} 

Further insight about the coupling between the FM and the NM layers  can be
obtained from the layer-dependent QDOS. In Fig.~4 the QDOS of a 1/3/1 sandwich
structure is shown for $T=0.1 T_C^0$ and three different values of the band
occupation of the NM layers ($n_{NM}=0.4,\,0.8,\,1.4$). For $n_{NM}=0.8$
the temperature dependence of the QDOS is plotted in Fig.~5 for
$T=0.1,\,0.9,\,1.0 T_C$. Two kinds of splittings are observed in the FM
spectrum. The strong Coulomb interaction between the electrons leads to
a splitting of the spectrum into a high and a low energy part (``Hubbard
splitting''). These two quasiparticle subbands (``Hubbard bands'') are
separated by an energy amount of the order $U$. In the lower subband
the electron mainly hops over empty lattice sites, whereas in the upper
subband it hops over lattice sites that are already occupied by another
electron with opposite spin. The weights of the subbands scale with the
probability that a propagating electron meets the one or the other
situation. The total weight of the QDOS of each layer is normalized to
1. In the strong coupling limit the weights of the lower and the upper
subbands are given by $(1-n_{\alpha -\sigma})$ and $n_{\alpha -\sigma}$,
respectively. Due to the vanishing Coulomb repulsion the Hubbard
splitting disappears in the NM spectrum. For temperatures below $T_C$,
an additional spin splitting (``exchange splitting'') in majority-
($\sigma=\uparrow$) and minority- ($\sigma=\downarrow$) spin direction
occurs in both the FM and the NM spectrum, leading to a non-zero
magnetization $m_\alpha=n_{\alpha\uparrow}-n_{\alpha\downarrow}$. Note
that, in principle, the NM and the FM spectrum for each spin direction 
occupy exactly the same energy region, thus preventing the electrons to be
trapped. However, the corresponding spectral weight may become very
small as can be seen in Fig.~4 and Fig.~5.

First we want to discuss the $n_{NM}$-dependence of the QDOS at low
temperatures. For the case of $d_{NM}=0$, the majority QDOS lies
completely below the chemical potential, and the system is fully
polarized ($n_{\alpha\uparrow}=1$) at $T=0$\cite{HPN98} (see also Fig.~1e).
If the NM overlayer is taken into account (Fig.~4), the majority QDOS of
both the interface and inner FM layers have tails above the chemical
potential due to the hybridization  between FM and NM layers. As a
consequence the system is no longer fully polarized at low temperatures
(see also Fig.~~1a,1b,1c,1d). The  weight of the tail determines the
reduction of the magnetization compared to the fully polarized state. 
Since the weight of the tail becomes smaller as the band occupation of
the NM layer is increased from $n_{NM}=0.4$ to $n_{NM}=1.4$ the
reduction of the magnetization in the interface gets weaker as well (see
also Fig.~~3c).  

Let us now turn to the NM layer. As mentioned above, the QDOS of the NM
layer has just one kind of splitting -- the exchange splitting. The QDOS
of the NM layer is quite different for majority- and minority- spin
direction. Because there is no Coulomb interaction in the NM layers, the
majority- and minority-QDOS do not affect each other. It is interesting
to note, that above the chemical potential the majority NM-QDOS
resembles the BDOS (Bloch density of states) of the square lattice which is
equivalent to the 
free standing fcc(100) monolayer. This is because for energies above the 
chemical potential a ($\sigma=\uparrow$)-electron within the NM layer is
effectively isolated since the spectral weight of the 
($\sigma=\uparrow$)-FM-QDOS is very small in this energy region.

For $n_{NM}=0.4$ (Fig.~~4a), there is very low spectral weight
in the majority NM-QDOS below the chemical potential.
As a consequence  the number of  spin-up
electrons is smaller than the number of spin-down electrons in the NM layer. 
The magnetization of the NM layer is negative, i.e., the NM and FM layers are
antiferromagnetically coupled.
With increasing $n_{NM}$ the center of gravity of the QDOS of the NM layer
gets shifted to lower  energies.   
When the peak of the majority NM-QDOS crosses the chemical potential, the 
number of the majority-spin electrons increases faster as a function of
$n_{NM}$ than the number of minority-spin  electrons. 
The magnetization of the NM layer increases and becomes
positive (see Fig.~~4b and 4c,also Fig.~~3b). Of course,  as $n_{NM}$
increases, the shape of 
the QDOS will also change. 
Until $n_{NM}=1.97$, the magnetization of the
NM layer will increase as a function of $n_{NM}$. Then it drops quickly,
because both the majority- and minority-spin QDOS gets shifted below the
chemical potential.

\begin{figure} 
\centerline{\epsfig{figure=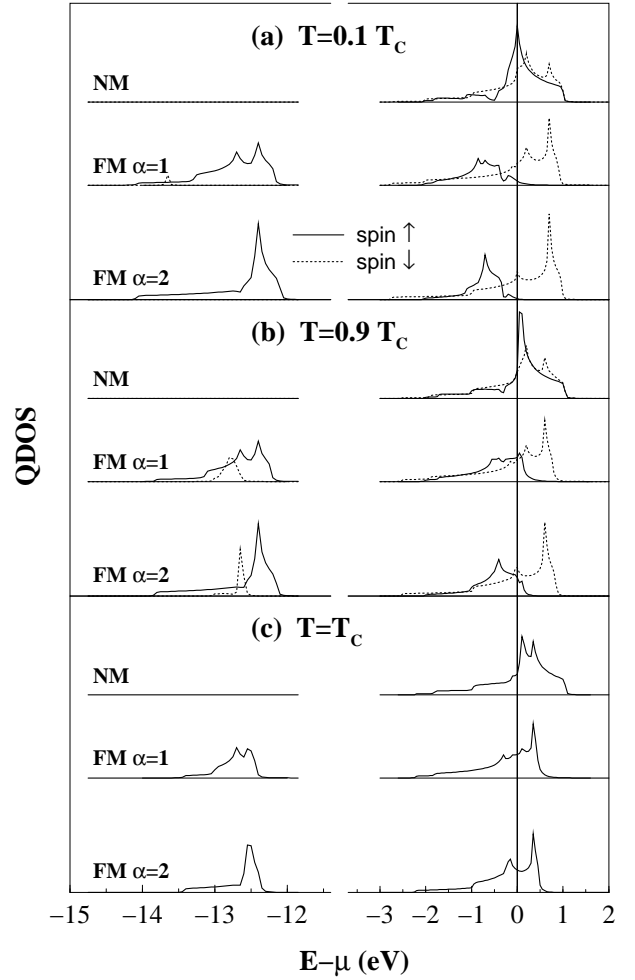,width=0.9\linewidth,angle=0}}
\vspace*{0.5cm}
\caption{Temperature dependence of the layer-dependent QDOS for a 1/3/1
  sandwich structure. (a) $T=0.1T_C^0$, (b) $T=0.9T_C$ and (c) $T=T_C$.}  
\end{figure} 

The temperature dependence of the FM-QDOS (Fig.~5) is dominated by two
correlation effects. As a function of increasing
temperature the spin-splitting between the centers of gravity of the majority
and minority quasiparticle subbands decreases. In addition there is a
temperature-dependent transfer of spectral weight between the lower and the
upper  quasiparticle subband.  The interplay of these two correlation effects
leads to a rather rapid demagnetization of the system as a function of
temperature and allows for 
Curie temperatures in a physically reasonable order of magnitude 
\cite{HPN98,NB89,HN97}. 
At $T=T_C$, the spin-splitting has
disappeared completely, both in the FM  and the NM spectrum (Fig.~~5c).

Due to the coupling between FM and NM layers, the NM-QDOS becomes temperature
dependent as well. While the band edges of the NM-QDOS stay fixed, there is a
redistribution of spectral weight within the NM-QDOS as a function of
temperature.   
We would like to point out that this temperature-dependence  
may even result in  a temperature induced  change from
ferromagnetic to antiferromagnetic coupling between FM and NM layers.
From Fig.~~5b, one can read off that at $T=0.9T_C$ the number of  spin-up
electrons is  smaller than the number of spin-down electrons.
Due to the reduced spin-splitting in the FM spectrum, the main peak of the
majority NM-QDOS  gets shifted above the chemical potential. 
As a consequence the coupling between the FM and the NM layer at
the interface  changes from ferromagnetic (Fig.~5a) to
antiferromagnetic (Fig.~5b) coupling. 
This kind of temperature induced change of the
coupling is observed only  for very thin NM layers ($d_{NM}=1$). 
For $d_{NM}>1$, we have not found such a behavior.

\begin{figure} 
\centerline{\epsfig{figure=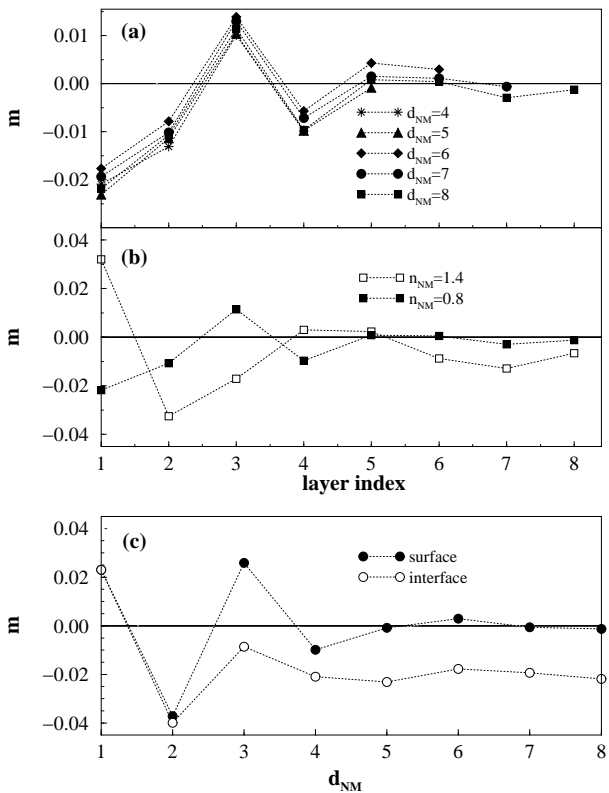,width=0.9\linewidth,angle=0}}
\vspace*{0.5cm}
\caption{The polarization of the NM layers for different numbers
  $d_{NM}$ of the NM layers. (a) Magnetizations of the NM layers as a
  function of the layer index,  (b) Magnetization profile of the NM
  layers for different band occupations of the NM layers. (c)
  Magnetizations of the NM interface and the surface layer   
  as a function of $d_{NM}$. Other parameters are: 
  $d_{FM}=4$, $T=0.1 T_C^B$.}  
\end{figure} 

Finally, we want to discuss the induced polarization in the NM layers which is 
shown in Fig.~6. The induced magnetization 
of the NM layers oscillates with the layer index for $d_{NM}\geq 4$.\cite{NOTE1}
 As a consequence, the total magnetization of the NM layers 
oscillates as a function of the thickness of the NM as well.
An  oscillatory behavior of total magnetization of the NM layers as a
function of the thickness of the NM has also been obtained by Bruno\cite{BRU95} 
within the theory of interlayer magnetic coupling. 
However, our starting  point and our results
are completely different. The quantum interference was
introduced to understand the properties of the NM layers. The
Bloch waves of the NM layers are regarded as to be confined within two FM
perturbations (or one FM perturbation and one Vacuum). Due to  the
constructive and destructive reflection at this confinement, the
densities of spin-up and/or spin-down electrons are found to be
oscillating as a function of the thickness of the NM. While the long
range magnetic order was excluded in his theory.
Within this concept, one can  find that the magnetization of the NM
interface should oscillate around zero. On the contrary, in our
calculation, the amplitude of the oscillation  decreases from the
interface to the surface (Fig.~~6a) and
the profile of the induced magnetization is almost independent of
$d_{NM}$(Fig.~~6a). The induced magnetization at the 
interface exhibits only a very small oscillation around a certain 
negative value as $d_{NM}\geq 4$ (see Fig.~6c). 
This indicates that the coupling between  FM
and NM layers is only affected by the properties of the FM and NM layers
close to the interface, whereas it is 
independent of the thickness of the NM overlayers as $d_{NM}\geq 4$. 

Due to the oscillatory behavior of the induced magnetization, 
we find the surface magnetization to be also oscillating 
as a function of the number of the NM layers (Fig.~6c). This surface
magnetization may be observed in experiment by use of spin polarized
photo-emission for very short mean free paths. We would like  to point
out that the oscillation of the NM magnetizations with respect to the
layer index is affected by the band occupation of the NM layers. The
amplitude as well as the  period of the oscillation changes when the
band occupation of the NM layers is changed (see Fig.~6b). 
 
\section{Conclusions} 

The effect of uncorrelated layers on the magnetic properties in thin
Hubbard films is studied by the spectral density approach (SDA). The
SDA, which reproduces the exact results of $t/U$-perturbation theory of
Harris and Lange\cite{HL67} in strong-coupling limit, leads to rather
convincing results concerning the magnetic properties of the Hubbard
model\cite{HPN98,HN99,NB89,HN97,PN96}. By comparison with different
approximation 
schemes for the Hubbard model as well as numerically exact QMC
calculation in the limit of infinite dimensions it has been
shown\cite{PHWN98} that the correct inclusion of the exact results of Harris
and Lange in the strong coupling regime \cite{HL67} is vital for a
reasonable description of the magnetic behavior of the Hubbard model,
especially at finite temperatures.  

Within our theory, FM and NM layers are treated on the same footing. 
The effects of the coupling between FM and NM on the magnetic properties
of the FM layers, as well as on the NM layers, have been studied in
detail. The Curie temperature of thin FM films is modified by the
presence of NM layers. For $n_{FM}=1.4$ and $n_{NM}=0.8$ the Curie
temperature increases gradually as a function of the number of FM
layers and converges to the corresponding bulk value for $d_{FM}=8-10$.
The induced polarization in the NM layers displays a long-range decreasing
oscillation with respect to the layer index of the NM layers. The
induced magnetization of the interface NM layer hardly changes as a 
function of number of the NM layers when $d_{NM}\geq 4$. 
This means the coupling between the FM and NM layers is determined by
the intrinsic properties of the FM and NM layers, such as the band
occupation, the in-site Coulomb interaction and the hopping, and it is
independent of the numbers of the FM and NM layers\cite{NOTE}. 

The magnetic properties of this thin film system have been
microscopically understood by means of spin- layer- and temperature-
dependent quasiparticle density of states (QDOS) for a 1/3/1 sandwich
structure. There exist two correlation induced 
band splittings in the FM spectrum. Besides the Hubbard splitting there
is an additional exchange splitting for temperatures below $T_C$. Due to
the vanishing Coulomb repulsion the Hubbard splitting disappears in the
NM spectrum. For the case of $d_{NM}=0$, the majority FM-QDOS lies
completely below the chemical potential, and the system is fully
polarized ($n_{\alpha\uparrow}=1$) at zero temperature\cite{HPN98}. If the
NM layers is taken into account, the majority QDOS of both the interface
and inner FM layers have tails above the chemical potential due to the
hybridization between the FM and NM layers. As a consequence the system
is no longer fully polarized at low temperatures. There is a reduction
of the FM magnetization compared to the fully polarized state. The
reduction of the magnetization of the interface FM layer gets weaker
as the band occupation of the NM layer is increased from $n_{NM}=0.4$ to 
$n_{NM}=1.4$. In addition for $n_{NM}=0.4$, there is very low spectral
weight in the majority NM-QDOS below the chemical potential. As a
consequence the number of spin-up electrons is smaller than the number
of spin-down electrons in the NM layer. The magnetization of the NM
layer is negative, i.e., the NM and FM layers are antiferromagnetic
coupled. As $n_{NM}$ increases, the magnetization of the NM layer
increases and becomes positive. The change of the coupling between
the FM and NM layers can also be induced by the temperature. This
temperature induced change of the coupling is observed only for very
thin NM layers. \\

\section*{Acknowledgements}

One of us (J.H.W.) wishes to acknowledge the Humboldt-Universit\"at for
the hospitality and financial support. Parts of this work have been done 
within the
Sonderforschungsbereich 290 (``Metallische d\"unne Filme: Struktur,
Magnetismus und elektronische Eigerschaften'') of the Deutsche
Forschungsgemeinschaft.


\begin{references}
\bibitem{NOTE1} In order to see the oscillation clearly, we relabel the layer
index, i.e., 1 refers to the interface, 2 next to the interface, etc.
\bibitem{NOTE}  We also calculated the induced polarization of the NM
  layers for different $d_{FM}$, and there was hardly change to be found for
  $d_{FM}>2$ .
\end{references}
\end{document}